\newcommand{\myname}{Shiqin Liu, Carl Higgs, Jonathan Arundel, Geoff Boeing, Nicholas Cerdera, David Moctezuma, Ester Cerin, Deepti Adlakha, Melanie Lowe, and Billie Giles-Corti}
\newcommand{\mynameabbrev}{S. Liu, C. Higgs, J. Arundel, G. Boeing, N. Cerdera, D. Moctezuma, E. Cerin, D. Adlakha, M. Lowe, and B. Giles-Corti}
\newcommand{\paperdate}{May 2021}
\newcommand{\papertitle}{A Generalized Framework for Measuring Pedestrian Accessibility around the World Using Open Data}
\newcommand{\papercitation}{\mynameabbrev. 2021. \papertitle. \textit{Geographical Analysis}, published online ahead of print. \href{https://doi.org/10.1111/gean.12290}{doi:10.1111/gean.12290}}
\newcommand{\paperkeywords}{Accessibility, Walkability, Open data, OpenStreetMap, Street network, Spatial analysis, Open source software}

\RequirePackage[l2tabu,orthodox]{nag}   
\documentclass[11pt,letterpaper]{article} 

\usepackage[T1]{fontenc}                
\usepackage[utf8]{inputenc}             
\usepackage{crimson}                    
\usepackage{helvet}                     

\usepackage[strict,autostyle]{csquotes} 
\usepackage[USenglish]{babel}           
\usepackage{microtype}                  

\usepackage{abstract}                   

\usepackage{authblk}                    
\usepackage{booktabs}                   
\usepackage{caption}                    
\usepackage[final]{draftwatermark}      
\usepackage{endnotes}                   
\usepackage{geometry}                   
\usepackage{graphicx}                   
\usepackage{hyperref}                   
\usepackage{natbib}                     
\usepackage{rotating}                   
\usepackage{setspace}                   
\usepackage{titlesec}                   
\usepackage{todonotes}
\usepackage{url}                        
\usepackage{amsmath}                    
\usepackage{appendix}                   

\graphicspath{{./figures/}}

\titleformat{\section}{\normalfont\sffamily\Large\bfseries\color{black}}{\thesection.}{0.3em}{}
\titleformat{\subsection}{\normalfont\sffamily\small\bfseries\color{black}}{\thesubsection.}{0.3em}{}
\titleformat{\subsubsection}{\normalfont\sffamily\small\color{black}}{\thesubsubsection.}{0.3em}{}

\hypersetup{
    pdfauthor={\myname},
    pdftitle={\papertitle},
    pdfsubject={\papertitle},
    pdfkeywords={\paperkeywords},
    pdffitwindow=true,         
    breaklinks=true,           
    colorlinks=false,          
    pdfborder={0 0 0}          
}

\SetWatermarkText{DRAFT}
\SetWatermarkScale{1.3}
\SetWatermarkLightness{0.9}

\begin{document}
    
    \title{\papertitle\footnote{{Citation info: \papercitation}}}
    \author[]{\myname}
    \date{\paperdate}
    \maketitle
    
\begin{abstract}
Pedestrian accessibility is an important factor in urban transport and land use policy and critical for creating healthy, sustainable cities. Developing and evaluating indicators measuring inequalities in pedestrian accessibility can help planners and policymakers benchmark and monitor the progress of city planning interventions. However, measuring and assessing indicators of urban design and transport features at high resolution worldwide to enable city comparisons is challenging due to limited availability of official, high quality, and comparable spatial data, as well as spatial analysis tools offering customizable frameworks for indicator construction and analysis. To address these challenges, this study develops an open source software framework to construct pedestrian accessibility indicators for cities using open and consistent data. It presents a generalized method to consistently measure pedestrian accessibility at high resolution and spatially aggregated scale, to allow for both within- and between-city analyses. The open source and open data methods developed in this study can be extended to other cities worldwide to support local planning and policymaking. The software is made publicly available for reuse in an open repository.
\end{abstract}

\section{Introduction}

Pedestrian accessibility refers to the extent to which the built environment supports walking access to destinations of interest. Monitoring spatial indicators of pedestrian accessibility helps planners and policymakers evaluate the impacts of urban design and transport interventions and guides targeted interventions towards creating healthy, sustainable cities, and achieving the United Nations (UN) Sustainable Development Goals (SDGs) \citep{united_nations_general_assembly_transforming_2015, united_new_urban_agenda_2016,organization_global_2016}. However, measuring spatial indicators of built environment features around the world to enable cities to be compared is challenging because local official data are either unavailable, not accessible to the public, or vary widely in coverage and format. Subject to data availability, indicators can be defined at very different scales of spatial aggregation by different government agencies, and the quality and usage of these indicators can be affected by local contexts. Indicators developed with traditional GIS software and non-comparable types of data often lack reproducibility and replicability \citep{brunsdon_opening_2020}. Existing spatial analysis tools can have limited functionality and transparency, or be expensive to use. These limit the ability of policymakers and researchers in resource-constrained contexts to derive indicators, reproduce results, or monitor change over time, further contributing to inequities in the ability to develop evidence-informed policy. 

To address these challenges, this study develops a framework using open source software and open data to deliver a generalized method for measuring pedestrian accessibility indicators within and between cities in diverse contexts. The method developed in this research enables calculation of key indicators of pedestrian accessibility, including but not limited to street connectivity, population density, daily living amenities access, and overall walkability. The open data approach allows researchers and practitioners to obtain relatively consistent data across cities and use them to calculate, analyze, and compare the indicators at both the local neighborhood scale and the spatially aggregated city level. This study also presents a validation framework to assess the quality of OpenStreetMap data and the suitability of the method. All analyses use the Python programming language and the software is publicly available for reuse in an open repository (https://github.com/global-healthy-liveable-cities/global-indicators), which can be customized to perform analyses for any city or study region. This can empower researchers, governments, and planning firms to evaluate spatial indicators and track ongoing progress of city planning and policy interventions.

This article is organized into five sections. First, the background and motivation for this research are presented. The next section introduces the open data sources used in this study for generating the indicators. The third section explains the methods of data processing and the indicator calculation workflow. The fourth section details the software usage. The article concludes by discussing implications for future research.

\section{Background}

\subsection{Challenge of assessing indicators of urban design and transport features}

Cities across the globe are experiencing increasing automobile dependency and urban sprawl, contributing to increasing traffic fatalities \citep{ewing_measuring_2003}, air pollution and environmental degradation \citep{wilson2013environmental}, climate change \citep{bart_urban_2010}, social segregation \citep{zhao_impact_2013}, and physical inactivity and obesity \citep{ewing_relationship_2008}. This threatens planetary and human health, quality of life, and wellbeing \citep{okeke_cities_2020, flannery_weather_2006, giles-corti_city_2016}. A growing body of research has investigated these issues and advocated for mitigation through pedestrian-friendly urban design and transit-oriented development \citep{boarnet_declining_2013,badami_urban_2009, ding_built_2012}. Designing pedestrian-friendly cities that promote active lifestyles will produce co-benefits for health and the environment. In the \textit{2030 Agenda for Sustainable Development} and the associated New Urban Agenda \citep{united_nations_general_assembly_transforming_2015, united_new_urban_agenda_2016}, the UN urged cities to provide more accessible, well-connected infrastructure for bringing people into public spaces and improving walkability. The World Health Organization (WHO) proposed improving people's access to basic amenities, including public transport, recreational spaces, markets, and public facilities as an essential strategy for promoting urban health equity \citep{who_commission_2008}. As specific urban design and transport features are needed to support healthier and more sustainable cities, developing and evaluating relevant indicators are crucial to monitor the ongoing progress for cities around the world toward achieving the UN SDGs and health equity goals in WHO's Global Action Plan \citep{giles-corti_city_2016, giles-corti_achieving_2020,badland_urban_2014, lowe_planning_2015}.

In general, comparing cities in diverse contexts worldwide using common indicators is challenging because the performance of indicators can be influenced by specific local contexts, such as climate, local culture and history, and development status. Measuring spatial indicators for cities around the world presents additional challenges, such as variations in spatial scale, geographical extent, measurement, and data availability and quality. First, indicators can be measured at different spatial aggregation scales, for example, for neighborhoods, census districts, municipalities, or the broader metropolitan expanse. Different tiers of government or institutional research groups may maintain their own indicator databases to address specific policy needs in certain geographical areas. Yet these indicators defined at various scales may have limited scope to support comparisons across study sites. The International Organization for Standardization (ISO) has published methodologies for calculating city service and quality of life indicators using consistent standards enabling cities to be compared \citep{iso2018}. However, these indicators report aggregated values at the city scale as a whole, which fails to provide essential information for identification of localized inequalities within cities \citep{giles-corti_achieving_2020}. 

Second, the geographical extent of a particular city or region often varies according to different authorities and interpretations. For example, the boundary of Melbourne, Australia may represent the 37 square kilometers of densely built up local government area, or the 9,993 square kilometers sprawling expanse of Australia's second most populous city \citep{melbourne_nodate}. Novel projects, such as the Global Human Settlements (GHS) Urban Centres Database (UCDB), seek to establish a consistent, shared geographic definition of \enquote{urban centres} globally \citep{ghs_ucdb_2015}. However, cities' administrative boundaries may be delineated by local authorities as either larger or smaller than their identified urban extents, and some cities may not be present in the GHS UCDB at all (e.g. Vic, Spain). Such inconsistencies in study region extent can bias the results of comparative studies across multiple study regions, particularly in international studies.

Third, a significant body of research is concerned with measurements of built environment exposures, many of which use GIS-based approaches and are conducted within a single region or across small samples of cities or countries at the macro level \citep{sarmiento_quality_2010,gomez_characteristics_2010, cambra_pedestrian_2012, marshall_community_2014}. However, the usage of heterogeneous methods and study region definitions means that meaningful comparison of indicators and the results of their analysis across studies and regions of interest is limited. This can explain inconsistent findings across studies in the associations between built environment characteristics and outcomes of interest, such as population health \citep{adams_international_2014}. The IPEN Adult Study \citep{sallis_physical_2016} assessed built environment attributes in relation to physical activity within and across 14 cities worldwide using proprietary GIS software with data sourced from government records and local survey samples, with local teams using a shared protocol for analysis. However, using a traditional desktop GIS-based approach to derive spatial indicators can be expensive, time-consuming and difficult to replicate and reproduce. Even though standards and detailed instructions for GIS methods can be established and documented to improve reproducibility and reliability of spatial work \citep{forsyth_standards_2006}, the inherent problems with expensive proprietary software offering limited \enquote{black box} sets of algorithms and source code versioning, means that they lack both the flexibility and rigour required for applied and reproducible research. 

Finally, spatial accessibility indicators require fine-grained street network and points-of-interest data that are traditionally sourced from official data repositories. However, local official spatial data can be difficult to obtain because they are either unavailable, inaccessible to the public, incomplete, out-of-date, or vary in coverage and format \citep{elwood2008grassroots, humphreys2013alcohol}. In addition, when working across different study region contexts, available data that may be considered at face value to represent comparable entities may not have been composed according to consistent criteria or classification principles. This compromises researchers' and practitioners' ability to calculate accurate, comparable, and reproducible built environment exposure measures across diverse study contexts.

Recent work has taken steps to address these challenges, but there remains a limited understanding of how policy settings and pedestrian accessibility relate to one another across cities globally. \citet{giles-corti_city_2016} proposed a framework of indicators to measure progress towards creating healthy and sustainable cities, including indicators of pedestrian accessibility. These researchers created a working group of healthy cities and GIS experts (i.e., the Global Healthy and Sustainable City Indicator Study Collaboration) to tackle this knowledge gap, adopting a federated model comprising a core team working together with geographically distributed collaborators across 25 cities in 19 countries. Each city has local representatives with expertise in healthy, active cities, policy context and spatial data who provided the local knowledge needed to undertake a comparative analysis of pedestrian accessibility and policy. A central team of healthy cities experts has developed a framework for assessment, coordinated activities, and analyzed patterns and trends across the individual city results. Mirroring this structure, the central GIS team developed the methods and tools to measure the policy-relevant spatial indicators of pedestrian accessibility outlined in this paper.

\subsection{Measuring pedestrian accessibility with open source software and open data}

The ability to measure and compare pedestrian accessibility indicators across cities can be limited by the challenges discussed above. But emerging \enquote{open} practices (e.g. open data, open source software) in spatial data science have brought new solutions to these problems. Availability of open data has made possible more fine-grained analyses for comparison of spatial features. One leading example is OpenStreetMap (OSM), the largest volunteered geographic information (VGI) online platform providing publicly accessible global spatial data \citep{OpenStreetMap}. This crowdsourced database, which is being compiled and updated continuously by volunteer users, has become increasingly popular in geography, transportation, planning and public health research as an open-source set of richly geographical and built environment data with global coverage. Many studies have suggested that OSM street network coverage is relatively complete in cities worldwide \citep{haklay_how_2010, barrington-leigh_worlds_2017}. Using OSM data, \cite{boeing_urban_2019} and \cite{boeing_street_2020} modeled and analyzed street networks around the world to better understand local and regional urban patterns. \cite{barrington-leigh_global_2019} assessed urban street connectivity and other walkability attributes to study the pattern of street-network sprawl within and across countries. OSM's growing popularity has also led to concerns about the quality and representation of its various geospatial features. The impact of contributor bias on OSM data coverage and completeness has also been noted by other researchers \citep{das2019gendered, quattrone2015there, basiri2019crowdsourced}. Acknowledging and addressing potential data issues are important to further unpack the value of this crowd-sources database for its wider use in planning and research communities.

Several open source tools for street and transportation network modelling using OSM data are relevant for accessibility analysis. For example, OSMnx, an open source Python-based software for modelling street network data based on OSM data \citep{boeing_osmnx_2017}, has been used for retrieval and analysis of spatial network features and indicators \citep{boeing_multi-scale_2020,natera_orozco_quantifying_2020,wang_road_2020, verendel_measuring_2019, calafiore_geographic_2021}. UrbanAccess and Pandana are Python software packages that can perform regional accessibility analyses using pedestrian and transit network data \citep{blanchard_urbanaccess_2017, foti_behavioral_2014}. Open source software has many advantages over proprietary software as it is publicly accessible, transparent, reusable, and customizable \citep{rey_show_2009}. Open source practices have grown in popularity in research as they improve reproducibility and replicability and bring new opportunities for collaboration \citep{boeing_right_2020}.

Nevertheless, comprehensive open source solutions for calculating spatial indicators of pedestrian accessibility worldwide remains limited. Walk Score \citep{Walk_Score_2020} offers a free tier of their API that can be used to retrieve up to 5,000 walk score index to measure neighborhood walkability for distinct addresses, but only in a few high-income countries including Australia, Canada, New Zealand and the United States. A more global city-focused approach was developed by the Institute for Transportation and Development Policy (ITDP) to measure within-city walkability indicators to produce a map database for GHS UCDB urban centres, which are available under Open Database Licence (ODbL) terms and can be visualized using an online interactive tool \citep{ITDP_pedestrians_nodate}. These projects have laid the groundwork to measure important aspects of walkability and accessibility in cities, highlighting the value of developing these indicators to guide city development.

Pedestrian accessibility plays a critical role in the creation of healthy and sustainable communities. \cite{giles-corti_city_2016} argued that cities need relevant indicators to provide pedestrian-friendly environments with highly connected streets, context-specific high density development, local public facilities, and pedestrian access to a mix of daily living destinations. While there are many determinants of health and physical activity, pedestrian accessibility enables healthy and sustainable lifestyles. Thus, quantifying pedestrian accessibility with spatial indicators is important to assess the extent to which a city's built environment will support healthy lifestyles.

\subsection{Current work}

Our work expands on recent research trajectories by developing a rigorous method for measuring pedestrian accessibility that can support urban policymakers around the world and in diverse contexts to address both within- and between-city inequities. As demonstrated in this study, quantifying pedestrian accessibility can be useful for calculating and bench-marking indicators for city planning that may encourage walking over motorised transport. To address data quality questions, methods for OSM data validation have been developed to investigate real-world representation and suitability for indicator calculation. We describe the method in this article using a case study project involving analysis of the 25 cities\footnote{The list of 25 cities is: Maiduguri, Nigeria; Mexico City, Mexico; Baltimore, United States; Phoenix, United States; Seattle, United States; Sao Paulo, Brazil; Hong Kong, China; Chennai, India; Bangkok, Thailand; Hanoi, Vietnam; Adelaide, Australia; Melbourne, Australia; Sydney, Australia; Auckland, New Zealand; Graz, Austria; Ghent, Belgium; Olomouc, Czech Republic; Odense, Denmark; Cologne, Germany; Lisbon, Portugal; Barcelona, Spain; Valencia, Spain; Vic, Spain; Bern, Switzerland; Belfast, United Kingdom} within the collaboration network developed by the core team. These 25 cities, spread across 19 lower-middle- to high-income countries, represent a diverse range of urban contexts. We worked with local GIS experts to assist with validation of the OSM data and assumptions used to construct the spatial indicators. In the present paper, we incorporate the case study to demonstrate how the method and software can be applied, validated, and generalized to calculate meaningful spatial indicators of pedestrian accessibility for cities across countries in diverse contexts.

\section{Data sources}

This study identifies relatively consistent open data sources that can be used for calculating the pedestrian accessibility indicators. We describe the normalized datasets derived from these open data sources that can be used as input to the main analysis for our case study cities using our software implementation. 

\subsection{Global Human Settlements Layer (GHSL)}

The GHSL is a repository with global scope produced by the European Commission using mixed data sources including census data, satellite imagery, and volunteered geographic information. The GHSL datasets are provided under CC BY 4.0 licence terms, and include vector representations of urban centres (e.g. Urban Centres Database, UCDB R2019A v1.2\footnote{UCDB R2019A: \url{https://jeodpp.jrc.ec.europa.eu/ftp/jrc-opendata/GHSL/GHS\_STAT\_UCDB2015MT\_GLOBE\_R2019A/}}; \citealt{ghs_ucdb_2015}), and population distribution raster data (e.g. GHS-POP R2019A\footnote{GHS-POP R2019A: \url{https://cidportal.jrc.ec.europa.eu/ftp/jrc-opendata/GHSL/GHS\_POP\_MT\_GLOBE\_R2019A/}}; \citealt{ghs_pop_2015}) at a series of epoch time points (1975, 1990, 2000 and 2015) and at high resolution. These datasets contain spatial and summary covariates which may be used to determine regions of interest and variation in population density or other attributes within an urban area. We used these GHSL datasets to derive urban boundaries and population for each region in the case study.

\subsection{OpenStreetMap}

OSM is a publicly accessible crowdsourced mapping platform with an open data ethos, with more than six million registered users since launching in 2004 \citep{OpenStreetMap}. Map data are provided through OSM under ODbL licence terms, collected by contributors based on various datasources such as ground surveys, satellite images, commercial and government data. OSM data can be used to construct the pedestrian street network and to derive datasets of points of interest (POIs) within a geographical region. Spatial features are tagged according to guidelines established over time by the OSM community using key-value pairs, the usage frequency and distribution of which may be queried using OSM Taginfo\footnote{OSM TagInfo: \url{https://taginfo.openstreetmap.org/}}. OSM is an important source for consistently coded road network data globally, with estimated completeness of coverage being very high for urban areas, and comparing favourably with equivalent official road datasets where these are available \citep{barrington-leigh_global_2019, zielstra_assessing_2013, zielstra_comparative_2011}. OSM has been used by researchers for tasks including deriving OSM points of interest data for studying neighborhood change \citep{zhang_using_2019} and analysis of food venue accessibility \citep{quinn_openstreetmap_2016, bright_openstreetmap_2018}. For our case study, we draw upon data from an OSM planet archive file\footnote{OSM planet archive: \url{https://planet.osm.org/pbf/planet-200803.osm.pbf}} dated 3 August 2020.

\subsection{Public transportation schedule data}

The General Transit Feed Specification (GTFS)
provides a standard framework for public transit agencies to disseminate public transportation timetable data and associated geographic information. Since its first creation by TriMet---a public transit agency in Portland, Oregon---and Google in 2005, GTFS has become the most popular transit service data format and has been used in various applications for planning transit trips, measuring accessibility, and creating timetables \citep{kujala_collection_2018, verbavatz_access_2020}. GTFS datasets vary in implementation by different agencies in different contexts, but usually comprise a set of text files containing comma-delimited data detailing transit agencies, a calendar schedule, stop locations, stop times, trips and routes data. Collectively, these data can be used to reconstruct and query a public transport schedule. GTFS data can be sourced directly from public transit agency websites, or an online open repository such as OpenMobilityData, subject to data availability. 

For our case study, GTFS data were audited for quality and downloaded for each of our study regions of interest targeting the year 2019, with approximate coverage of April to May in the northern hemisphere, and October to December in the southern hemisphere. These time windows are intended to capture the school term before summer school holidays for seasonal consistency between cities, as weather and other factors may influence transport schedules and behavior.

\section{Methods}

\begin{figure}[htbp]
    \centering
    \includegraphics[width=0.9\textwidth]{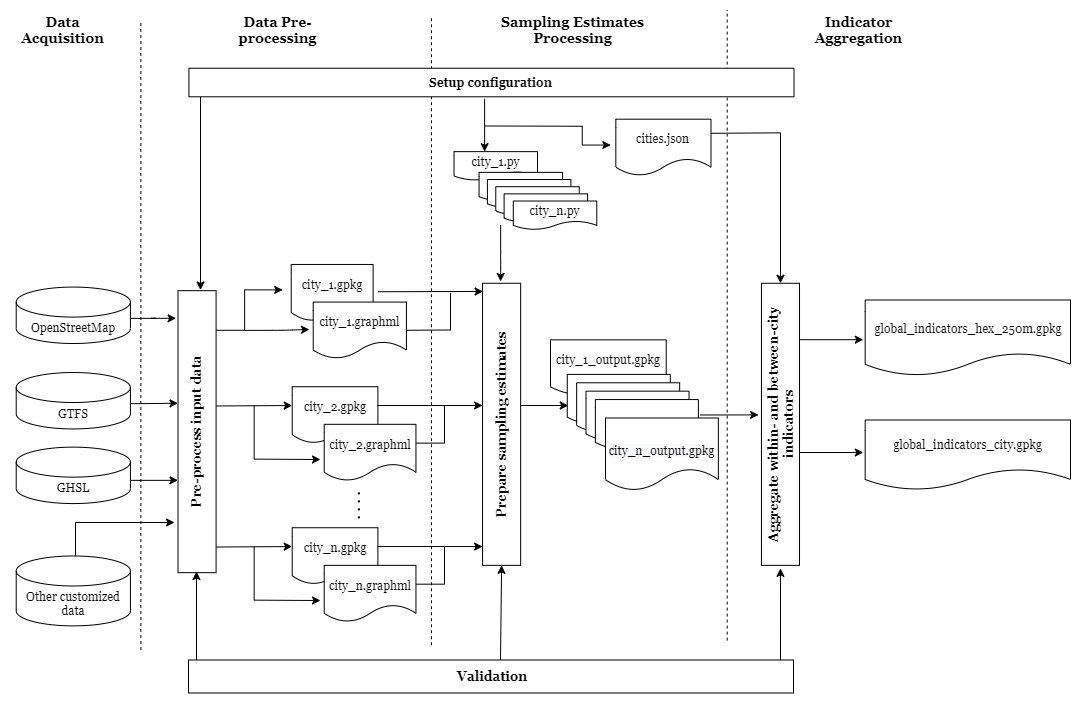}
    \caption{Indicator calculation workflow}
    \label{fig:process_figure}
\end{figure}

We used these source data in a generalized Python-based framework to measure pedestrian accessibility consistently across diverse urban study regions at three configurable scales: fully disaggregate sample points at regular intervals along the pedestrian network of populated urban regions (in our case study, 30 m, serving as proxies for residential address locations), aggregate small area hexagon cells (our case study used a 250 m diagonal width grid, approximating our population source data resolution), and aggregate city-level summaries. This section presents the five major components of our methodological framework: 1) project and study region parameters configuration, 2) input data pre-processing, 3) sampling estimates processing, 4) indicator aggregation, 5) validation. The process flowchart diagram appears in Figure \ref{fig:process_figure}.

\subsection{Project and study region parameter configuration}

Project and study region specific parameters are defined external to the code using configuration files. These files detail aspects include: default settings for analyses, such as the study region buffer distance, sampling frequency, accessibility evaluation distance and local neighborhood analysis distance; the location of key input resources in the project data folder, such as the GHS UCDB, GHS-POP, an OSM derived dataset; and parameters to contextualize distinct study regions. This external configuration framework means that users can adjust these parameters to incorporate different study regions and data sources, or re-parameterize for a completely new project with distinct requirements and outcomes of interest. Examples for these files and parameters were set up for the case study cities and can be found in our open repository\footnote{Default parameters are chosen for our case study. Project and region-specific parameters are configured in this script: \url{https://github.com/global-healthy-liveable-cities/global-indicators/blob/main/process/setup_config.py}.}.

\subsection{Input data pre-processing}

Data pre-processing for each study region is undertaken through processing a series of Python scripts drawing upon project and region parameters to populate a Docker-ized Postgres PostGIS spatial database of intermediary resources, which can be drawn upon to prepare layers within study region-specific GeoPackages used as input to the main analysis workflow.

\subsubsection{Urban boundaries}

Study region boundaries are defined using a flexible combination of official administrative boundaries and the UCDB. Based on the data availability and advice gathered from local experts, we designate three scenarios to justify the urban extent of study regions: 1) where both datasets are available in a study region, the intersection between the UCDB and official boundary feature would be used; 2) if the UCDB urban feature is available and validated to best represent a study region of interest, this would be used directly; 3) if the UCDB urban feature is not available for a study region of interest, a bespoke vector boundary would be used. 

To account for edge effects when undertaking neighborhood accessibility analyses, a study region buffer is defined in order to retrieve spatial data sources used for analysis up to an appropriate distance beyond the edge of the study region (e.g. 1600 m, which is further than the accessibility analyses undertaken in our case study). In addition to the overall study region boundaries, a tesselated grid of hexagonal polygons is generated in order to summarize the spatial distribution of local results at a meaningful scale (250 m in our case study) for each study region. This makes use of an open source hex grid function \citep{Hugh_hex-grid_2015} for PostGIS. 

\subsubsection{Pedestrian street network}

Street network data are retrieved from OSM via the Overpass API for buffered boundaries of study regions using the Python package OSMnx, and used to derive a network appropriate for pedestrian walking behaviours in addition to a cleaned dataset of intersections for evaluation of street connectivity \citep{boeing_osmnx_2017}. Network resources are exported as a graphml file for use in subsequent network analyses, but also imported as edge and node layers into the study region spatial database for use in further pre-processing.

\subsubsection{Population data and densities}

Population data are used to account for local neighborhood population densities, evaluate the spatial distribution of population density across study regions, and estimate the percentage of population with access to destinations. Our process generates a virtual raster table using the global set of GHS-POP image data, from which a subset of the global population intersecting each study region's buffered boundary is extracted and saved. Population estimates are derived for each hexagonal grid cell as the average of intersecting population raster cells, and for the overall study region as the sum of intersecting population raster cells. Population and street intersection density estimates for hexagonal grid and study region boundaries are calculated by dividing the respective count estimates by the associated polygon areas in square kilometers. 

\subsubsection{Points of interest (POIs)}

Points of interest are obtained from OSM to calculate pedestrian access to destinations. A polygon filter file is generated for each study region's buffered boundary and used to extract a subset of relevant OSM data defined in the configuration stage; this is ingested into each study region's database as tables of point, line and polygon features with geometry transformed to match locale-specific projected coordinate reference systems. A dataset of POIs classified by destination types is created for each study region within the buffered boundary; where identified features of interest have geometries which are polygons (e.g. building footprints) or lines, these are represented instead using their centroid points.

In our case study, we consulted the OSM tagging guidelines along with OSM TagInfo to identify sets of appropriate key-value pair tags which we used to define different destination types of interest, for example, food markets and supermarkets, convenience stores, and public transport locations. 

\subsubsection{Sample points}

For an analysis concerned with population accessibility, point locations for residential addresses may be the ideal sampling units for use in origin-destination (O-D) network analyses. However, address point data can be difficult to acquire, and where available, they rarely represent actual residences or dwellings. Therefore, we develop a generalized sampling approach for constructing O-D metrics to measure within-city local neighborhood accessibility. Sample points are generated every 30 meters along the derived pedestrian street network within each study region, and are filtered to those located within hexagonal cells meeting a configurable minimum population threshold criterion. In our case study focusing on urban contexts, located within grid cells with an estimated population of 5 or more were retained, representing a likelihood that at least one household resides in this small area; in this way, our sample points serve as proxy locations for residential dwellings. For each sample point, we calculate and store as attributes the respective distances to the two terminal nodes of the street network edge along which it is generated. This information is used to facilitate a two step analysis for deriving sample point indicator estimates, detailed in the following section.

\subsection{Sample point estimates processing}

Local neighborhood estimates of population, street intersection densities and destination accessibility are evaluated using a multi-step process: a first pass analysis is undertaken using node points from the routable pedestrian network; a second pass analysis is then conducted to derive estimates for the full set of sample point locations within each city. Sample points can be considered as representative of local address points. The output estimates at the sample point level from this implementation process includes population density, street intersection density, and accessibility to points of interest. In addition, we configure the process to evaluate the composite scores for access to daily living destinations and walkability.

\begin{figure}[htbp]
    \centering
    \includegraphics[width=0.75\textwidth]{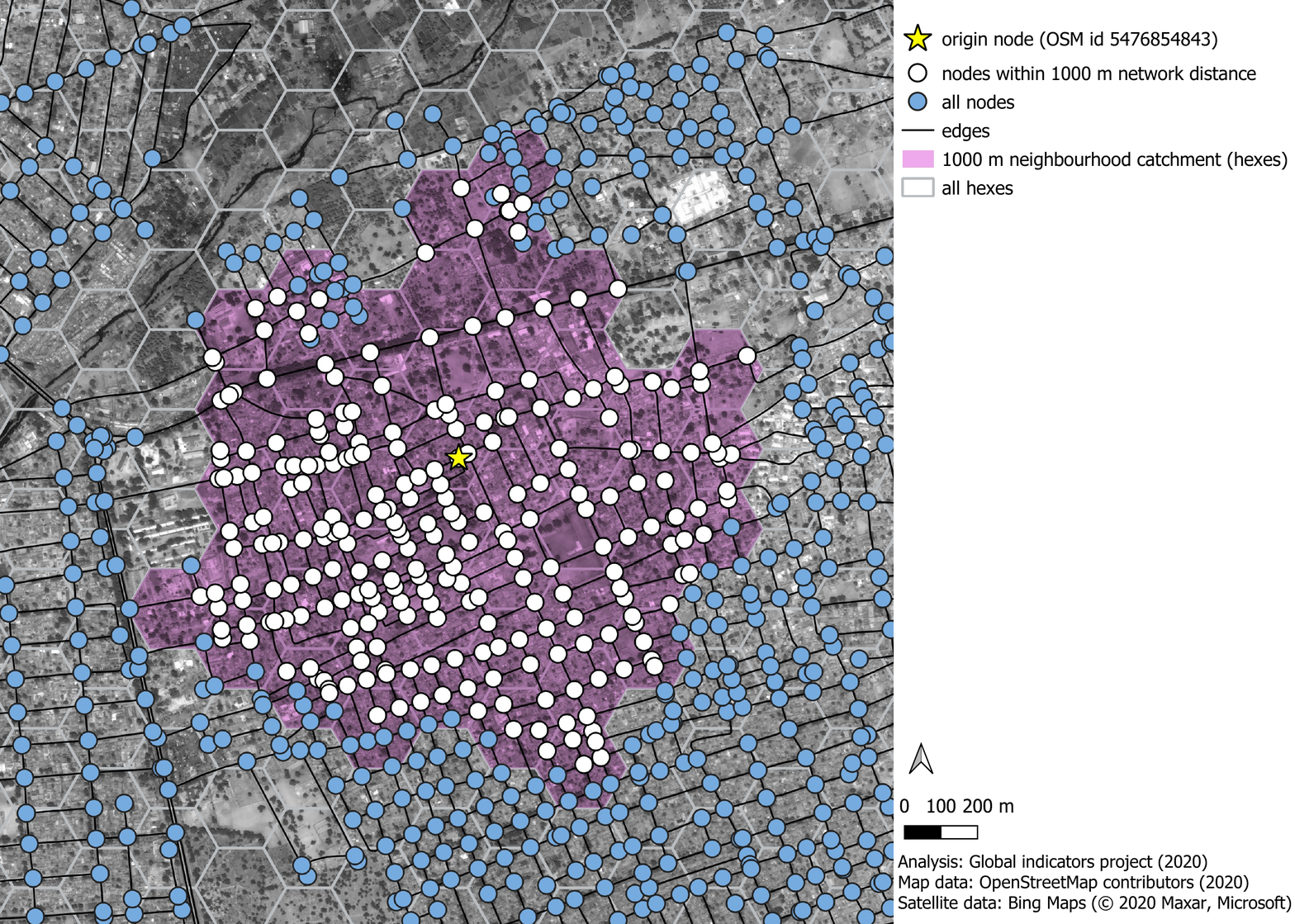}
    \caption{Example map illustrating calculation of the local walkable neighborhood area for one node.}
    \label{fig:local_nh_nodes}
\end{figure}

For neighborhood analysis of network nodes, each node is associated with the unique identifier of the small area hexagonal grid cell it is located within using a spatial join. The all-pairs weighted Dijkstra algorithm is used to identify for each node all other nodes reachable within a distance threshold (e.g. 1000 m in our case study, representing a locally walkable catchment). This information is used to associate each node with the set of hexagonal grid cells comprising its local walkable neighborhood. This approach approximates the \enquote{sausage buffer} technique for deriving a buffered walkable neighborhood expanse used in comparable studies of local neighbourhood walkability \citep{forsyth_sausage_2012,higgs_urban_2019}. The average of the small area densities associated with these cells produce local neighborhood population and street intersection density estimates for each node. Dijkstra's shortest path algorithm for local neighborhood processing proved orders of magnitude faster than generating subgraphs over large geographical extent of node. This process is illustrated in Figure \ref{fig:local_nh_nodes}.

\begin{figure}[htbp]
    \centering
    \includegraphics[width=0.75\textwidth]{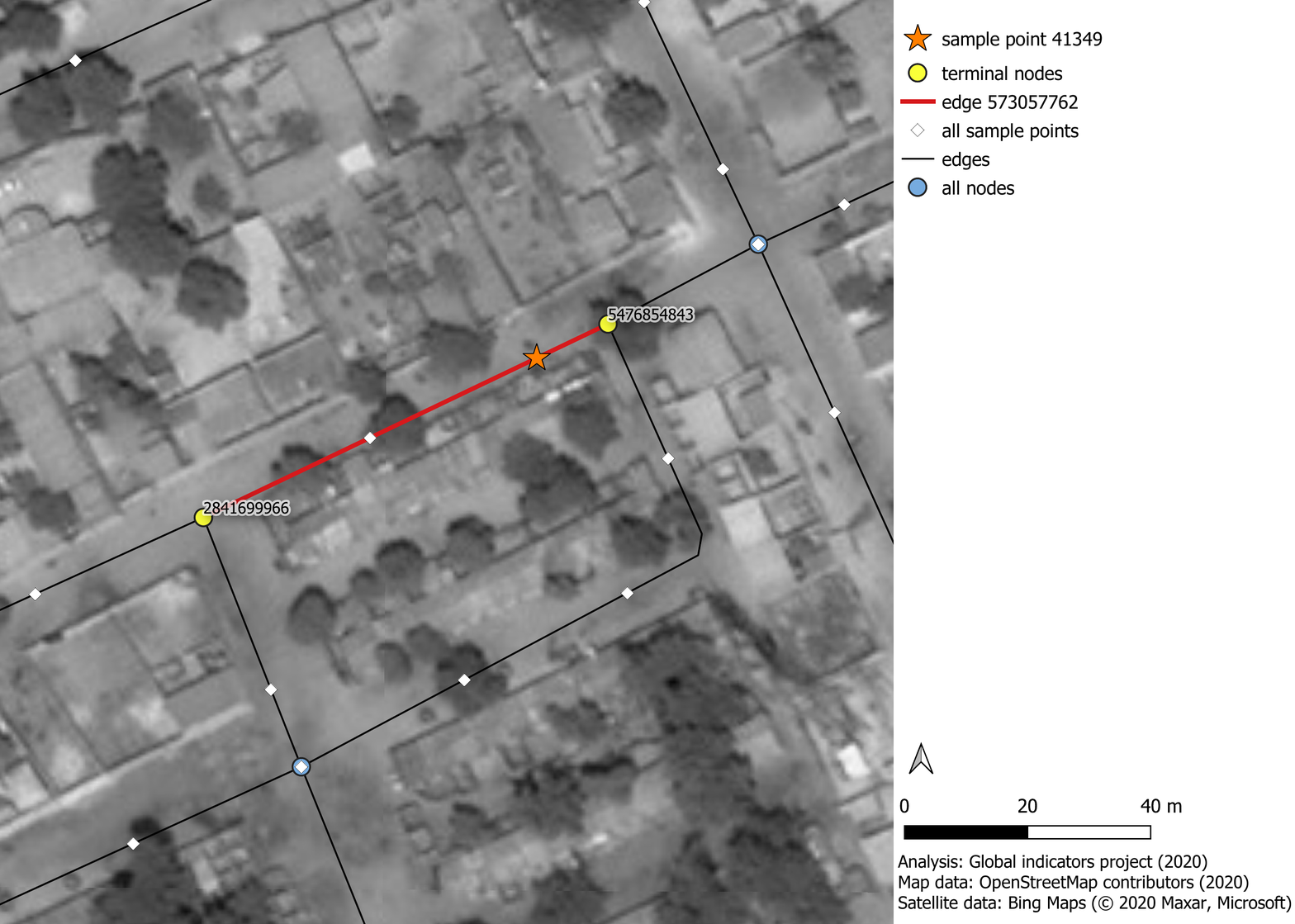}
    \caption{Example map demonstrating how each sample point is pre-associated with the unique identifiers of the nodes located at the terminal ends of the OSM edge segment.}
    \label{fig:local_nh_sample_point}
\end{figure}

Sample point density estimates are derived through association with node-level results based on sample points' distance relationships with their associated terminal nodes. Using an undirected pedestrian network, each of $j$ sample points $s_j$ is either associated with two nodes accessible by some distance at either end of the edge segment on which they are located, or associated twice with a single node, in the case of a small cul-de-sac. These relations with $i$ nodes $n_{s_j}^i$ are depicted in Figure \ref{fig:local_nh_sample_point}.  In the edge case where either one of the two terminal node unique identifiers recorded for each sample point is not present in the nodes dataset, the sample point would be omitted (this could occur in rare instances near the study region boundary, where stretches of road terminate beyond the study region buffer). Broadly, node relationships can be distinguished by sample points being either coincident with a terminal node (located at 0 m distance from it; for example, generated at the origin of an edge linestring), or some positive distance from both terminal nodes.

To derive the $k$th density estimate for the $j$th sample point ($\hat{d}_{s_j}^k$), if that point is coincident with one of its two associated nodes (ie. because distance $l_{(s_j,n_{s_j}^i)}$ is zero) then it is directly associated with that node's density estimates $d_{n_{s_j}^i}^k$, since this measurement is directly representative of the sample point's location.  For the remaining sample points, the node distance relationships are evaluated such that the $k$th density estimate for the $j$th sample point is estimated using the average weighted by the complement of nodes' relative distances (ie. node density estimates are weighted commensurate to sample point proximity).

\[ 
\hat{d}_{s_j}^k = \begin{cases}
    d_{n_{s_j}^1}^k,& \text{if } l_{(s_j,n_{s_j}^1)} = 0\\
    d_{n_{s_j}^2}^k,& \text{if } l_{(s_j,n_{s_j}^2)} = 0\\
    \sum_{i=1}^{n=2} \left( 1-\frac{l_{(s_j,n_{s_j}^i)}}{l_{(s_j,n_{s_j}^1)} + l_{(s_j,n_{s_j}^2)}} \right) d_{n_{s_j}^i}^k, & \text{otherwise} 
    \end{cases}
\]

Destination accessibility can be measured by first evaluating the shortest distance that could be travelled between an origin and destination, and then determining whether this meets some criteria for access. Our tool incorporates accessibility functions to perform shortest path and aggregation queries for origins and destinations using a pedestrian routable network of nodes and edges, tuned for performance at scale. Derived pedestrian network nodes and edges are loaded into Pandana, and the distance from each node to the closest destination within a given search distance threshold of interest is calculated. The distance for evaluation of access is a configurable parameter in our software. The distance of 500 meters may correspond approximately to a 5-to-10-minute walk, and policy recommendations for access to public transport and services in different countries often use a similar value as a policy target \citep{organization_global_2016, united_nations_general_assembly_transforming_2015}.

Then, we evaluate the distance from a sample point to its two associated nodes, and derive the minimum full distance from a sample point to its nearest destination. The full distance of a sample point is then converted to an accessibility score with regard to a target threshold. In our framework, accessibility scores can be calculated using binary or gravity-based continuous measures. Specifically, we provide three optional functions following existing studies \citep{higgs_urban_2019, vale_active_2016, vale_influence_2017, yang_walking_2012} (See Table \ref{tab:access_table}).
In our case study, we employed a binary score to evaluate destination accessibility (1 if access the estimated access distance is within 500 m, and 0 otherwise). The binary access measure is a preferable choice in our case study for its ease of implementation and interpretation, and this method may be less stringent than the distance-decay based measure considering the imperfect representation of the open data. We used this accessibility method to generate sample point estimates of accessibility to points of interest such as fresh food outlets, convenience, and public transport stop locations.  

\begin{table}[htbp]
    \centering
    \caption{Type of Pedestrian Accessibility Measures}
    \label{tab:access_table}
    \renewcommand{\arraystretch}{1} 
    \resizebox{\textwidth}{!}{%
        \begin{tabular}{lll}
            \hline
            \textbf{Type of measures}    & \textbf{Functions} & \textbf{Default parameters} \\ \hline
            \textbf{Binary}              &  $ A{i} =   \left\{\begin{matrix} 1, & if\ d_i{}_j <= t \\  0, & if\ d_i{}_j > t \end{matrix}\right.$  & t=500m \\ \hline
            \textbf{Soft threshold}      & $A{i} = \frac{1}{1+\exp^{k\frac{(d_i{}_j-t)}{t}}}$ & t=500m, k=5 \\ \hline
            \textbf{Cumulative-gaussian} &   $A{i} = \left\{\begin{matrix} 1, & if\ d_i{}_j <= t \\  \exp^{-\frac{{(d_i{}_j-t)^2}}{v}}, & if\ d_i{}_j > t \end{matrix}\right.$ & t=500m, v=129842 \\ \hline
            \multicolumn{3}{l}{\begin{tabular}[c]{@{}l@{}}Note: $A{i}$ is the access score for point i, $d_i{}_j$ is the pedestrian network distance from point i to nearest destination j.\\ t is the threshold distance parameter, and k, v are parameters for the slope of decay. These parameters are user-configurable. \\ The default parameters are set up for each function as a suggestion.  \\ References: \cite{higgs_urban_2019, vale_active_2016, vale_influence_2017, yang_walking_2012} \end{tabular}}
        \end{tabular}%
    }
\end{table}

The daily living score is a composite measure of access to daily living destinations within recommended distances, which has been used as a readily calculable proxy for land use mix and diversity that supports walkability \citep{badland_identifying_2017}. It is calculated by aggregating the sum of binary accessibility scores from sample points to three core kinds daily living destinations (i.e. fresh food outlets, convenience, and public transport). This results in a local neighborhood daily living score for a particular location's access to diverse destinations, ranging from zero to three.  

Walkability for each sample point is calculated as the sum of standardized z-scores of local walkable neighborhood street intersection and population densities and the daily living destinations score \citep{higgs_urban_2019}. 

Estimates for the above measures can be generated for sample points within a city using the pre-processed input data through modification of a configuration file and command-line execution of a Python script in our tool, with results saved as a new layer in the city-specific sample point output GeoPackage. 

\subsection{Indicator aggregation}

After sample point estimates are calculated, a separate aggregation process is conducted to summarize them into relevant indicators for within- and between-city comparisons. The aggregation process includes three major components: a) calculate the average of sample point estimates for urban grids within each study region; b) calculate standardized z-scores of indicators relative to all study regions; c) aggregate city average and population weighted average of indicators. This process can be implemented by running a single Python script in our presented tool.

The indicator results are generated at two distinct geographical scales: fine-grained hexagonal grid level and city level. Table \ref{tab:table1} shows the examples of fine-grained indicator attributes and descriptions, and these results are saved in a city layer within an output GeoPackage file (i.e. global\_indicators\_hex\_250m.gpkg). Similarly, Table \ref{tab:table2} presents the examples of aggregated city level indicator attributes that are saved in a separate output GeoPackage (i.e. global\_indicators\_city.gpkg). These attributes include urban covariates such as area, population density, street intersection density, and number of sample points in a study region; within-city indicators such as spatial and population weighted accessibility to various destinations, daily living score and walkability; and between-city indicators in standardized scores to facilitate comparison.

\begin{table}[htbp]
    \centering
    \caption{Example of Grid-level Output Indicators in global\_indicators\_hex\_250m.gpkg}
    \label{tab:table1}
    \resizebox{\textwidth}{!}{%
        \begin{tabular}{ll}
            \toprule
            Indicator  & Description \\
            \toprule
            \multicolumn{2}{l}{\textbf{Urban covariates}} \\
            \midrule
            Study region & Study region name \\
            Population per sqkm	& Population per square kilometer of urban study region \\
            Intersections	& Street intersection count \\ 
            Intersections per sqkm	& Street intersections per square kilometer of urban study region \\
            urban\_sample\_point\_count	& Number of sample points generated along derived pedestrian network \\
            \toprule
            \multicolumn{2}{l}{\textbf{Within city, percentage of sample point with access within 500 meters walking distance to…}} \\
            \midrule
            pct\_access\_500m\_fresh\_food\_market\_binary	& a supermarket \\
            pct\_access\_500m\_convenience\_binary	& a convenience store \\
            pct\_access\_500m\_pt\_any\_binary	& any public transport \\
            pct\_access\_500m\_public\_open\_space\_any\_binary	& any public open space \\
            \toprule
            \multicolumn{2}{l}{\textbf{Within city, spatial average …}} \\
            \midrule
            local\_nh\_population\_density	& local neighborhood population per square kilometer \\
            local\_nh\_intersection\_density & local neighbourhood intersections per square kilometer \\
            local\_daily\_living	& daily living score \\
            local\_walkability	& walkability index (sum of z-scores of population density, intersection density and daily living score) \\
            \toprule
            \multicolumn{2}{l}{\textbf{Between city, spatial average …}} \\
            \midrule
            all\_cities\_z\_nh\_population\_density	& local neighborhood population per square kilometer (z-score relative to all cities) \\
            all\_cities\_z\_nh\_intersection\_density	& local neighborhood intersections per square kilometer (z-score relative to all cities) \\
            all\_cities\_z\_daily\_living	& daily living score (z-score relative to all cities) \\
            all\_cities\_walkability & walkability index (sum of z-scores relative to all cities) \\
            \bottomrule
        \end{tabular}%
    }
\end{table}

\begin{table}[htbp]
    \centering
    \caption{Example of City-level Output Indicators in global\_indicators\_city.gpkg}
    \label{tab:table2}
    \resizebox{\textwidth}{!}{%
        \begin{tabular}{ll}
            \toprule
            Indicator  & Description \\
            \toprule
            \multicolumn{2}{l}{\textbf{Urban covariates}} \\
            \midrule
            Study region & Study region name \\
            Area (sqkm) & Urban study region area (square kilometer) \\
            Population estimate & Urban study   region population estimate \\
            Population per sqkm & Population per square kilometer of urban study region \\
            Intersections & Street intersection count \\
            Intersections per sqkm & Street intersections per square kilometer of urban study region \\
            urban\_sample\_point\_count & Number of sample points generated along derived pedestrian network \\
            \toprule
            \multicolumn{2}{l}{\textbf{Within city,   percentage of population with access within 500 meters walking distance to…}} \\
            \midrule
            pop\_pct\_access\_500m\_fresh\_food\_market\_binary & a supermarket \\
            pop\_pct\_access\_500m\_convenience\_binary & a convenience   store \\
            pop\_pct\_access\_500m\_pt\_any\_binary & any public transport \\
            pop\_pct\_access\_500m\_public\_open\_space\_any\_binary & any public open   space \\
            
            \toprule
            \multicolumn{2}{l}{\textbf{Within city, population weighted average…}} \\
            \midrule
            pop\_nh\_pop\_density & local   neighborhood population per square kilometer \\
            pop\_nh\_intersection\_density & local neighborhood street intersections per square kilometer \\
            pop\_daily\_living & daily living   score (access to supermarket, convenience, and public transport) \\
            pop\_walkability & walkability index (sum of z-scores of population density, intersection density and daily living score) \\
            \toprule
            \multicolumn{2}{l}{\textbf{Between   city, population weighted average…}} \\
            \midrule
            all\_cities\_pop\_z\_nh\_population\_density & local   neighborhood population per square kilometer (z-score relative to all   cities) \\
            all\_cities\_pop\_z\_nh\_intersection\_density & local neighborhood street intersections per square kilometer (z-score relative to all   cities) \\
            all\_cities\_pop\_z\_daily\_living & daily living   score (z-score relative to all cities) \\
            all\_cities\_pop\_walkability & walkability   index (sum of z-scores relative to all cities) \\
            \bottomrule
        \end{tabular}%
    }
\end{table}

\subsection{Validation}

To assess the quality of the OSM data, suitability and external validity of the open data method, we develop the following validation framework and present validation findings for our case study cities. 

\subsubsection{Local partner validation}

Local partner validation includes a qualitative analysis to assess the extent to which identified open data aligns with the local context of a study region and to examine the face validity of the indicators. In this process, a descriptive report of the preliminary assets is prepared for each study region following pre-processing detailing the identified urban area's boundaries and distribution of amenities of interest, and disseminated for review by collaborators with local area knowledge. Feedback on data accuracy, completeness and suggestions for improvement is solicited through a survey for each study region. This allows researchers to make further revision and improvement on data and method as required based on feedback received. 

For our case study, we conducted validation with local partners from the 25 included cities. Feedback received from local collaborators was used to iteratively refine boundary definitions and key-value pairs used to identify points of interest, to help ensure fair representation was made of diverse local contexts. For example, in consultation with local knowledge, the study region boundary for Mexico City was expanded beyond the federal district to include the metropolitan area of the Valley of Mexico, and the OSM key-value pairs were expanded to capture more local fresh food outlets. Through this process, we also found a lack of amenities on OSM for Maiduguri, Nigeria, and thus a custom dataset of amenities of interest was collected by collaborators for the analysis. Following calculation of built environment indicators of accessibility, spatial distribution maps were disseminated and reviewed to confirm expectations of subject matter experts familiar with the local contexts. Four waves of qualitative validation were conducted throughout the course of the project regarding the spatial and network analyses for their cities to ensure the methods and inputs used met expectations of local subject matter experts: 1) preliminary feasibility scoping; 2) data availability (e.g. providing official data sets for specific features of interest); 3) preliminary data validation; 4) spatial distribution of indicator validation.

\subsubsection{OSM destination and edge validation}

OSM destination and edge validation comprises a quantitative analysis to compare OSM derived data against available local official data for evaluating the data quality, representation, and suitability for indicator calculation. In this process, the proportion of overlap between the OSM derived dataset (i.e. edges and destinations) and official dataset within 10 and 50-meter buffer are first evaluated, in line with other studies that address road networks' representation with specific thresholds \citep{sehra_extending_2020}. Second, a suitability analysis is conducted to evaluate the accessibility method (i.e., hexagon grid validation), which does not distinguish between hexagon neighborhoods that have one destination or ten destinations: in either scenario, there is access to at least one destination. In this step, an additional hexagonal grid layer is overlaid onto the edge layer to identify the distribution of destination points located across hexagons. This validation identifies two conditions: a \enquote{true} condition is assigned in a hexagon if both OSM and official destination points are present or if both are not present; and a \enquote{false} condition is assigned if a destination point is present in only one but not the other. Following this, a weight is given on each hexagon according to the percentages of OSM and official destination points: this is calculated by dividing the identified points from each dataset by the total sum of destination points inside each hexagon. The following summary statistics are produced to justify the spatial representation and difference of OSM and official destinations across hexagons: 1) the proportion of hexagons with a \enquote{true} condition; 2) the average weight of OSM and official destination points, relative to the total amount of hexagons; and 3) the average weight of OSM and official destination points, relative to the total amount of hexagons with a \enquote{true} condition. 

To implement this step of the validation, official datasets for street network and relevant destination data were solicited from local collaborators, following the local partner validation, for four of our case study cities: Belfast, Hong Kong, Olomouc, and Sao Paulo. Official edge data were collected from Belfast, Hong Kong, and Olomouc. For all three cities, more than 90\% of the edges in the official dataset fell within 10 meters of an edge from the OSM derived dataset (see Table \ref{tab:table3}). This suggested that the official street network edges were either a) accounted for in the OSM derived dataset; b) relatively close to edges from the OSM derived dataset. Additionally, when comparing the total length, the OSM derived datasets were about twice as long as their official counterparts. Generally, the segments of the official datasets that were not covered by OSM occured in areas with high OSM coverage. For each of the three cities that provided data, more than 98\% of edges from the official dataset intersect with edged derived from OSM derived edges that are buffered by 50 meters. Most of the missing edge segments appear to occur under specific circumstances like planned developments and access roads on private property. These areas would not be counted as a part of the pedestrian network and thereby would not affect the results of the walkability analysis. 

\begin{table}[htbp]
    \centering
    \caption{Edge Validation Indicator Table}
    \label{tab:table3}
    \resizebox{\textwidth}{!}{%
        \begin{tabular}{lllll}
            \hline
            \multicolumn{1}{|c|}{{\color[HTML]{24292E} \textbf{City}}} & \multicolumn{1}{c|}{{\color[HTML]{24292E} \textbf{\begin{tabular}[c]{@{}c@{}}Total Length of \\ OSM Edges (m)\end{tabular}}}} & \multicolumn{1}{c|}{{\color[HTML]{24292E} \textbf{\begin{tabular}[c]{@{}c@{}}Total Length of \\ Official Edges (m)\end{tabular}}}} & \multicolumn{1}{c|}{{\color[HTML]{24292E} \textbf{\begin{tabular}[c]{@{}c@{}}Percent Length of Official Edges \\ that Overlap with Buffered OSM Edges \\ (buffer = 10m)\end{tabular}}}} & \multicolumn{1}{c|}{{\color[HTML]{24292E} \textbf{\begin{tabular}[c]{@{}c@{}}Percent Length of Official Edges \\ that Overlap with Buffered OSM Edges \\ (buffer = 50m)\end{tabular}}}} \\ \hline
            \multicolumn{1}{|l|}{{\color[HTML]{24292E} \textbf{Olomouc}}} & \multicolumn{1}{l|}{{\color[HTML]{24292E} 616000}} & \multicolumn{1}{l|}{{\color[HTML]{24292E} 310000}} & \multicolumn{1}{l|}{{\color[HTML]{24292E} 93.40\%}} & \multicolumn{1}{l|}{{\color[HTML]{24292E} 98.10\%}} \\ \hline
            \multicolumn{1}{|l|}{{\color[HTML]{24292E} \textbf{Belfast}}} & \multicolumn{1}{l|}{{\color[HTML]{24292E} 1700000}} & \multicolumn{1}{l|}{{\color[HTML]{24292E} 1330000}} & \multicolumn{1}{l|}{{\color[HTML]{24292E} 90.70\%}} & \multicolumn{1}{l|}{{\color[HTML]{24292E} 98.10\%}} \\ \hline
            \multicolumn{1}{|l|}{{\color[HTML]{24292E} \textbf{Hong Kong}}} & \multicolumn{1}{l|}{{\color[HTML]{24292E} 7217000}} & \multicolumn{1}{l|}{{\color[HTML]{24292E} 2911000}} & \multicolumn{1}{l|}{{\color[HTML]{24292E} 98.90\%}} & \multicolumn{1}{l|}{{\color[HTML]{24292E} 99.90\%}} \\ \hline
        \end{tabular}%
    }
\end{table}

Official destination data of fresh food locations were collected from Belfast, Olomouc, and Sao Paulo. But the official Belfast data were unsuitable for accessibility analyses because they were provided as a polygon of retail land use classification; this would not result in a fair comparison with OSM point destinations due to the mismatch in destination types and morphology, thus the Belfast data were excluded from the validation. For the other two cities, the results show that there was a relatively low overlap between the official and OSM derived destinations. Approximately 50\% of points intersected from one dataset to the other when buffered by 50 meters (see Table \ref{tab:table4}). Explanations for this low correspondence may include: a) each dataset could be a different snapshot in time of the constantly changing landscape of fresh food markets (e.g. stores may cease operations, leading to errors in both datasets); b) the official dataset could be vulnerable to becoming outdated due to a dependency on data collection practices maintained by the local governing body; c) there could be a poor semantic match of destination classification criteria (e.g. between an official dataset of retail POIs and a derived dataset of OSM fresh food markets; or a street market which occurs one day each week, but may not meet the OSM fresh food market tag query). 

For the hexagon grid validation, as shown in Table \ref{tab:table5}, more than 80\% of hexagons in both cities met the \enquote{true} condition (i.e., if both OSM and official destination points are present or if both are not present). The average weights of the OSM destinations were 0.41\% (relative to all hexagons), and 4.72\% (relative to hexagons with \enquote{true} condition) higher than that of the official destinations in the city of Olomouc. In the city of Sao Paulo, a wider discrepancy was present with the average weights of the OSM destinations 15.89\% (relative to all hexagons), and 4.80\% (relative to hexagons with \enquote{true} condition) higher than that of the official destinations. This is largely attributed to a small count of official destinations in Sao Paulo compared to the OSM derived destinations (see Table \ref{tab:table4}). Nonetheless, the average weights were consistently higher for the OSM derived destinations than the official one. Together with their overall high correspondence, we concluded that while the OSM data diverged from the official data in various ways, they were sufficient, or even better, for the purposes of calculating the accessibility indicators using our method. 

\begin{table}[htbp]
    \centering
    \caption{Destinations Validation Indicator Table}
    \label{tab:table4}
    \resizebox{\textwidth}{!}{%
        \begin{tabular}{lllllll}
            \hline
            \multicolumn{1}{|c|}{{\color[HTML]{24292E} \textbf{City}}} & \multicolumn{1}{c|}{{\color[HTML]{24292E} \textbf{\begin{tabular}[c]{@{}c@{}}Count of OSM \\ Destinations in the Core\end{tabular}}}} & \multicolumn{1}{c|}{{\color[HTML]{24292E} \textbf{\begin{tabular}[c]{@{}c@{}}Count of Official \\ Destinations in the Core\end{tabular}}}} & \multicolumn{1}{c|}{{\color[HTML]{24292E} \textbf{\begin{tabular}[c]{@{}c@{}}Percent of Buffered OSM \\ Destinations Intersect with \\ Buffered Official Destinations \\ (buffer = 10m)\end{tabular}}}} & \multicolumn{1}{c|}{{\color[HTML]{24292E} \textbf{\begin{tabular}[c]{@{}c@{}}Percent of Buffered Official \\ Destinations Intersect with \\ Buffered OSM Destinations \\ (buffer = 10m)\end{tabular}}}} & \multicolumn{1}{c|}{{\color[HTML]{24292E} \textbf{\begin{tabular}[c]{@{}c@{}}Percent of Buffered Official \\ Destinations Intersect with \\ Buffered Official Destinations \\ (buffer = 50m)\end{tabular}}}} & \multicolumn{1}{c|}{{\color[HTML]{24292E} \textbf{\begin{tabular}[c]{@{}c@{}}Percent of Buffered Official \\ Destinations Intersect with \\ Buffered OSM Destinations \\ (buffer = 50m)\end{tabular}}}} \\ \hline
            \multicolumn{1}{|l|}{{\color[HTML]{24292E} \textbf{Olomouc}}} & \multicolumn{1}{l|}{{\color[HTML]{24292E} 51}} & \multicolumn{1}{l|}{{\color[HTML]{24292E} 36}} & \multicolumn{1}{l|}{{\color[HTML]{24292E} 20.00\%}} & \multicolumn{1}{l|}{{\color[HTML]{24292E} 31.66\%}} & \multicolumn{1}{l|}{{\color[HTML]{24292E} 50.00\%}} & \multicolumn{1}{l|}{{\color[HTML]{24292E} 51.66\%}} \\ \hline
            \multicolumn{1}{|l|}{{\color[HTML]{24292E} \textbf{Sao Paulo}}} & \multicolumn{1}{l|}{{\color[HTML]{24292E} 797}} & \multicolumn{1}{l|}{{\color[HTML]{24292E} 12}} & \multicolumn{1}{l|}{{\color[HTML]{24292E} 1.72\%}} & \multicolumn{1}{l|}{{\color[HTML]{24292E} 0.70\%}} & \multicolumn{1}{l|}{{\color[HTML]{24292E} 13.50\%}} & \multicolumn{1}{l|}{{\color[HTML]{24292E} 1.08\%}} \\ \hline
        \end{tabular}%
    }
\end{table}

\begin{table}[htbp]
    \centering
    \caption{Hexagon Grid Validation Indicator Table}
    \label{tab:table5}
    \resizebox{\textwidth}{!}{%
        \begin{tabular}{|l|l|l|l|l|l|}
            \hline
            \multicolumn{1}{|c|}{{\color[HTML]{24292E} \textbf{City}}} &
            \multicolumn{1}{c|}{{\color[HTML]{24292E} \textbf{\begin{tabular}[c]{@{}c@{}}Proportion of Hexagons \\ with True Condition\end{tabular}}}} &
            \multicolumn{1}{c|}{{\color[HTML]{24292E} \textbf{\begin{tabular}[c]{@{}c@{}}Average Weight of \\ OSM Destinations, \\ relative to the Total \\ Amount of Hexagons\end{tabular}}}} &
            \multicolumn{1}{c|}{{\color[HTML]{24292E} \textbf{\begin{tabular}[c]{@{}c@{}}Average Weight of  \\ Official Destinations, \\ relative to Total \\ Amount of Hexagons\end{tabular}}}} &
            \multicolumn{1}{c|}{{\color[HTML]{24292E} \textbf{\begin{tabular}[c]{@{}c@{}}Average Weight of \\ OSM Destinations, \\ relative to total \\ Amount of Hexagons \\ with True Condition\end{tabular}}}} &
            \multicolumn{1}{c|}{{\color[HTML]{24292E} \textbf{\begin{tabular}[c]{@{}c@{}}Average Weight of \\ Official   Destinations, \\ relative to the Total \\ Amount of Hexagons \\ with True Condition\end{tabular}}}} \\ \hline
            {\color[HTML]{24292E} \textbf{Olomouc}} &
            {\color[HTML]{24292E} 89.69\%} &
            {\color[HTML]{24292E} 10.09\%} &
            {\color[HTML]{24292E} 10.50\%} &
            {\color[HTML]{24292E} 29.72\%} &
            {\color[HTML]{24292E} 25.00\%} \\ \hline
            {\color[HTML]{24292E} \textbf{Sao Paulo}} &
            {\color[HTML]{24292E} 83.68\%} &
            {\color[HTML]{24292E} 16.28\%} &
            {\color[HTML]{24292E} 0.39\%} &
            {\color[HTML]{24292E} 5.26\%} &
            {\color[HTML]{24292E} 0.46\%} \\ \hline
        \end{tabular}%
    }
\end{table}

\subsubsection{Virtual ground truth validation}

Virtual ground truth validation is conducted to further evaluate the extent to which the open data are aligned with reality. Specifically, this process is executed to understand the prevalence of false positive OSM derived destinations. This is done by comparing relevant destination locations to what exists on three Google services and assigning a \enquote{true} or \enquote{false} value for each: Google Maps View (tag of the location), Google Satellite View (building footprint), and Google Street View (Ground image). A sampling approach is employed to select destinations of interest for comparison. We first associate the population density hexagon with each destination. Then, we divide the destinations into five quintiles stratified by their respective population densities in each study region. For each quintile, ten destinations are selected at random, resulting in a total of fifty destinations for each individual city for the ground truth validation analysis. Upon comparison of all three Google services, each destination point is assigned a \enquote{true} value if two or more true instances are present; otherwise, a \enquote{false} value is assigned. When an image is not available on Google Street View, the value is determined by the Google Maps View and Google Satellite View.

We implemented the ground truth validation for all of the case study cities. The results indicated that an average of 87.6\% of the OSM-derived destinations corresponded with equivalent destinations on Google services. Additionally, 45\% of times when the destination did not appear on the Google Maps View, it was locateable using Google Street View. This suggested that cities with low street view coverage were likely to have higher than expected instances in which OSM destinations were not able to be matched with Google services. For example, Chennai and Hanoi both had an average of approximately 30\% of OSM destinations with a \enquote{false} value on available Google services; this in part can be explained by a high absence of an image on Google Street View for these cities. In Chennai and Hannoi, 90\% and 70\% of sampled destination locations did not have street view imagery, respectively. In the absence of street view imagery, the limited validation was conducted using Google Maps View and Google Satellite View. With the exception of these two cities, all other cities in our case study demonstrated high correspondence between the sampled OSM-derived destinations and destinations on Google services.

\section{Software reuse}

\subsection{Software usage and code reusability}

The methods described above have been delivered as an open source software tool that is freely accessible from a public GitHub repository that can be run using a custom Docker container environment. This allows users to reuse the code and implement the analytical process within a stable computational environment. Detailed instructions for running code have been provided in the open repository to enable users with limited programming experience to employ the process. Furthermore, users can customize the process to employ additional customized data, indicator variables, study regions, and parameter configurations based on their own project needs through modification of the project configuration files. The generalized framework and the open data approach developed in this research can be applied to diverse urban contexts worldwide, and used by academics and practitioners to support planning and policymaking, public health, and urban development research. 

\subsection{Computational environment and requirements}

The Linux-based Docker environment contains all the Python packages required for implementing the analytical workflow, alongside an installation of LaTeX for generating relevant documentation and reports. The environment uses Python 3.7 and the Python geospatial data science stack, including \texttt{pandas}, \texttt{geopandas}, \texttt{pysal}, \texttt{scipy}, and \texttt{numpy}, as well as \texttt{NetworkX} and \texttt{OSMnx} for spatial graph modeling and analysis. The workflow's code has been optimized to run efficiently, even for large study sites at the scale of the urban area, on a modestly configured laptop (e.g., using an Intel i5 processor and 8 GB of RAM).

\section{Discussion and future directions}

This paper presented an open source software framework for processing and validating open data, and calculating pedestrian accessibility indicators for configurable study regions. The open data approach and the generalized framework enable calculation of a mix of pedestrian accessibility indicators from the street scale to the local neighborhood scale to the city scale. The framework helps address the challenges and limitations of measuring built environment features in the existing literature in terms of data availability, reproducibility, and generalizability. 
This software framework offers several advantages. First, it is open source. Second, it measures pedestrian accessibility concerning both population and spatial distribution across a diverse set of cities worldwide. Many existing studies or tools focus instead on spatial features of accessibility for either city-scale or within-city analysis. Third, our method allows indicators to be mapped and analyzed fully spatially disaggregated at the point level, and aggregated at both high resolution grid and city levels. This helps researchers and practitioners seeking to map and identify inequalities of access to services within and between cities, or use these local estimates as covariates in research involving statistical modelling and simulation. Fourth, the open data approach facilitates creation of relatively consistent measures for cities over large and diverse geographical extents. No data are perfect, but this effort is undertaken to identify the best available sources for consistent indicator analyses of urban environments, with a global scope. 

This framework has been validated with local knowledge from collaborators in our case study cities. The validation results demonstrate that neither official data nor alternative open data represent the real world perfectly. However, comparing the two offers an understanding of the relative completeness of the open data, and provides us with insight into how our methodological and data choices impact the results. While the open data display some divergence from approximately comparable official data, there is a relatively high correspondence such that local contexts are adequately represented in most instances. Thus, open data were found to be broadly suitable for the purposes of deriving indicators of pedestrian accessibility in diverse urban contexts. Our work highlights many strengths of using OSM over government and commercial data for spatial analysis, such as its free accessibility, broader study region coverage (e.g. not restricted to a municipal administrative boundary, or a central business district), more timely data representation, and greater capacity for customization with consistently query-able data which can be updated by the public to improve it.

Nevertheless, open data have important limitations. Our results from virtual ground truth validation show that destination amenities may not be well-captured in Google Street View imagery in certain cities (e.g. Chennai, Hannoi). This finding is consistent with the study by \cite{fry_assessing_2020}, which suggested that informal communities in the Global South are more likely to have an under-representation of Google Street View images. Our analyses across different study regions suggest that the detail, precision, and accuracy of OSM coverage varies across cities and countries, and some locations may lack representation of basic amenity features. The OSM community is not equally invested in rich and poor neighborhoods, and the data do not have equal capacity in urban and rural coverage, or developed and developing countries. \cite{barrington-leigh_worlds_2017} and \cite{haklay_how_2010} pointed out that rural or low-density communities are most likely to have missing streets than urban areas. Thus, indicators constructed from this framework, if solely relying on the OSM data, are likely to overestimate accessibility in high-capacity communities and underestimate them in low-capacity spaces. Our present framework identifies and addresses these potential biases through a combination of quantitative and qualitative data validation methods. To mitigate potential data issues, we suggest caution when using OSM data outside urban areas known to have adequate data quality and coverage. In addition, our framework offers the alternatives of using a combination of official and open data, or incorporating other customized data to achieve better and more reliable results of the indicators. For example, in addition to the customisation of destination data, our data pre-processing approach allows for the use of custom boundaries for study regions of interest. Our research highlights the potential to further unlock the value of the open data ecosystem through encouraging members in countries, where the data are poorly served, to routinely collect spatial data, and contribute to this crowd-sourced database.

By making our source code openly available with a stable working environment and documentation, researchers and practitioners are able to fully reproduce the analytical process with a few commands. The methods are scalable and customizable for the inclusion of alternative data sources and study regions, helping to remove barriers to application of indicators in policymaking in less-resourced settings. The indicator output generated from the tool can be used to compare and track within- and between-cities' inequities and planning progress and to support public health, urban development, and transportation research. Finally, this software has been used by the core research team to produce and analyze spatial indicators of urban design and public transport features across 25 cities worldwide. As it is an open source project, we welcome a community of researchers to participate in and benefit from its ongoing development, improvement, and extension.

\setlength{\bibsep}{0.00cm plus 0.05cm} 
\bibliographystyle{apalike}
\bibliography{references}

\end{document}